\documentclass[twocolumn,preprintnumbers,floatfix,prb,superscriptaddress]{revtex4}
\usepackage{graphicx}
\usepackage{graphicx}
\usepackage{dcolumn} 
\usepackage{bm}
\usepackage{epsfig}
\begin{document} 

\title{Evaluation of Compensated Magnetism in La$_2$VCuO$_6$: \\
           Exploration of Charge States}

\author{Victor Pardo}
 \email{vpardo@ucdavis.edu}
\affiliation{Department of Physics,
  University of California, Davis, CA 95616
}

\author{Warren E. Pickett}
 \email{wepickett@ucdavis.edu}
\affiliation{Department of Physics,
  University of California, Davis, CA 95616
}


\begin{abstract}

We present an ab initio study of double perovskite La$_2$VCuO$_6$, which was one of the earliest compounds suggested as a potential compensated half-metal. Two charge and spin configurations close in energy have been identified. (i) The originally envisioned spin-compensated V$^{4+}$:d$^1$/Cu$^{2+}$:d$^1$ configuration is comprised of antialigned S=1/2 cations. This state is a spin-compensated half-metal for moderate values of U (the on-site Coulomb repulsion strength on the metal ions) and is insulating for larger values of U. (ii) An unanticipated non-magnetic solution V$^{5+}$:d$^0$/Cu$^{+}$:d$^{10}$ consists of an empty- and a full-shell ion, both spherically symmetric, that leads to a band insulator. This ionic band insulator is calculated to be the ground state at small and moderate values of U. The different distortions of the perovskite structure that occur for each state are central in determining the energy differences. Treating the Cu$^{2+}$ Jahn-Teller distortion self-consistently is particularly important for the magnetic solution.

\end{abstract}

\maketitle


\section{Background}
Half-metals comprise an important class of materials that has been studied extensively in recent years.\cite{sr2femoo6,katsnelsonRMP,hmafm_review_jssc,hmafm_review_ange} They are particularly interesting because of their anticipated applications in the field of spintronics. This type of material is conducting in one spin channel, but is gapped in the other.  The idea of having such materials has been studied for nearly three decades.\cite{degroot,cro2_schwarz} This half-metallic property could eventually prove feasible for providing injection of 100\% spin-polarized carriers,which would provide a great enhancement in capabilities over current spintronics materials.  

In recent years, a twist of the argument has appeared by suggesting that one could have a half-metal with compensating moments, hence with the absence of macroscopic fields.  This state is produced with an equal number of electrons in both spin channels,\cite{pickett_hmafm} leading to no net moment, making the system robust against external magnetic fields. This situation can arise in magnetic materials with
antiferromagnetic (AF) interactions between the magnetic ions, but such that they lead to partially filled bands in one spin channel and an insulating gap in the other. This possibility has been explored ab initio extensively, and a few materials have been suggested to display this effect.\cite{pickett_hmafm,vpardo_hmaf,la2vcuo6_guo,androulakis,lasrvruo6,lacavmoo6,lacavoso6,hm_cpl} This property has been explored in perovskites,\cite{hmafm_per,hmafm_dp_prb,hmafm_dp_ssc,hmafm_dp_chen_jmmm,hmafm_dp_chen_jap} multilayers,\cite{hmafm_ml_physb,hmafm_ml_jmmm} pnictides,\cite{hmafm_pnic_jpcc,hmafm_pnic_jap,hmafm_pnic_jpcm} Heusler alloys\cite{hmafm_heus_jpsj,hmafm_heus_jpcm,hmafm_heus_jmmm,hmafm_heus_prb} and even diluted magnetic semiconductors.\cite{hmafm_dms} Interestingly, it has been suggested that this so-called spin-compensated half-metals eventually could lead to a new type of superconductivity.\cite{pickett_sss}

One of the studied compounds has been La$_2$VCuO$_6$,\cite{pickett_hmafm,vpardo_hmaf,la2vcuo6_guo} where it has been suggested as a possible spin-compensated half-metal,\cite{pickett_hmafm} but at large values of U it has been proposed to be an insulator.\cite{vpardo_hmaf} In this paper we explore new possible electronic phases in this material, apart from the previously studied spin-compensated solution, analyzing ab initio the corresponding lattice distortions that can occur and how they relate to each of the electronic phases considered.

\section{Computational methods and Structure}
Our electronic structure calculations were  performed within density functional
theory \cite{dft} using the all-electron, full potential code {\sc wien2k} \cite{wien}
based on the augmented plane wave plus local orbital (APW+lo) basis set.\cite{sjo}
The generalized gradient approximation\cite{gga} (GGA) was used for the calculations at U=0.
To deal with  strong correlation effects we apply the LDA+U
scheme\cite{sic1,sic2} including an on-site repulsion U and Hund's coupling J 
for the $3d$ states. We have explored various values of U for Cu and V, and we discuss below the different solutions that can be obtained, but our primary results presented here are obtained using U= 6 eV for Cu and U= 3.5 eV for V. For all the solutions presented, we have relaxed the internal positions and also the volume.


La$_2$VCuO$_6$ is studied in an ordered double perovskite structure. Due to the cation size mismatch between V and Cu and also due to the expected distortions of the octahedral cage, we have optimized the structure (volume and internal coordinates) in the tetragonal space group (I4/mmm, no. 139), allowing oxygen relaxations around the cations. Two inequivalent oxygens occur in the structure at positions 8h ($x$,$x$,0) and 4e (0,0,$z$).

\begin{table}[h!]
\caption{Summary of interatomic distances in the two structures that were considered: sc (spin-compensated) and nm (non-magnetic)}\label{tabdist}
\begin{center}
\begin{tabular}{|c|c|c|}
\hline
 & V-O (\AA) & Cu-O (\AA) \\
\hline
\hline
short sc & 2 $\times$ 1.81 & 4 $\times$ 1.94  \\
short nm & 4 $\times$ 1.83 & 2 $\times$ 2.01 \\
long sc  & 4 $\times$ 1.92  & 2 $\times$ 2.05  \\
long nm  & 2 $\times$ 1.86  & 4 $\times$ 2.03 \\
\hline
\end{tabular}
\end{center} 
\end{table}

\section{Discussion}

We have obtained two self-consistent solutions, one that we will call a spin-compensated state (which has been discussed in the past), but also an unexpected non-magnetic solution. The first one corresponds to a formal oxidation state of  Cu$^{2+}$:d$^9$, V$^{4+}$:d$^{1}$, with both atoms in a spin-half state. The magnetic coupling between the cations is AF, and a spin-compensated configuration results. The second is non-magnetic, with no unfilled shells, corresponding to a simple ionic Cu$^{+}$:d$^{10}$ and V$^{5+}$:d$^{0}$ compound. This emergence of high and low charge states with very large formal charge difference is unexpected, and becomes interesting as a non-magnetic competitor for the ground state.

The differing charge and magnetic states lead to different types of atomic relaxation. The cation-oxygen distances obtained from our structure optimizations are summarized in Table \ref{tabdist}. The spin-compensated state accommodates the V$^{4+}$ $d^1$ electron in a d$_{xy}$ orbital by stretching the oxygen octahedron in the $x-y$ plane, whereas the octahedron surrounding Cu is elongated along the $z$-axis to produce one hole in the $d_{x^2-y^2}$ orbital, which becomes raised in energy by such distortion. Apart from relaxing the internal oxygen coordinates, we have also optimized the unit cell volume for both states, leading to $a$= 3.87\AA~($c$/$a$= 1.00). The internal coordinates were relaxed within LDA+U, so the particular orbital configuration is obtained self-consistently. Within GGA alone, only a non-magnetic solution can be obtained; volume optimization in that case leads to $a$= 3.96\AA.

The difference in atomic radii between Cu and V leads to a larger Cu-O than V-O distance. The ionic radii for Cu$^{2+}$ and Cu$^+$ are, respectively, 0.87 \AA\ and 0.91 \AA. In the case of V, the ones we need to consider are 0.72 \AA\ and 0.68 \AA\ for V$^{4+}$ and V$^{5+}$, respectively. The elongation (respectively, shortening) along the c-axis of the oxygen octahedra around Cu (V) is of approximately 5 (6)\%. For the non-magnetic solution, the spherical symmetry of the ions leads to a much smaller distortion of the octahedral cages.

The structure proposed here differs with respect to a recent report,\cite{la2vcuo6_guo} where a half-metallic AF solution with spin-half cations has been obtained, similar to our spin-compensated solution but with smaller magnetic moments.  In that report a roughly undistorted octahedral cage surrounding both cations was obtained from a structural optimization, the metal-oxygen distance resulting the same along the three axes. In particular, no Jahn-Teller distortion was predicted. Our results show that a structure optimization with such electronic structure distorts considerably the octahedral environment due to a Jahn-Teller effect, once a hole occupies one of the e$_g$ orbitals of the Cu$^{2+}$ cations. According to our calculations, the solution proposed in Ref. \onlinecite{la2vcuo6_guo} would be unstable and eventually lead, when relaxed with U= 0, to a non-magnetic solution. For obtaining the spin-compensated solution we describe here, one needs to use U$\ne$ 0 during the optimization, but it is also necessary to start from a Jahn-Teller distorted environment for both the V and Cu surrounding oxygen octahedra.


\begin{figure}[ht]
\begin{center}
\includegraphics[width=\columnwidth,draft=false]{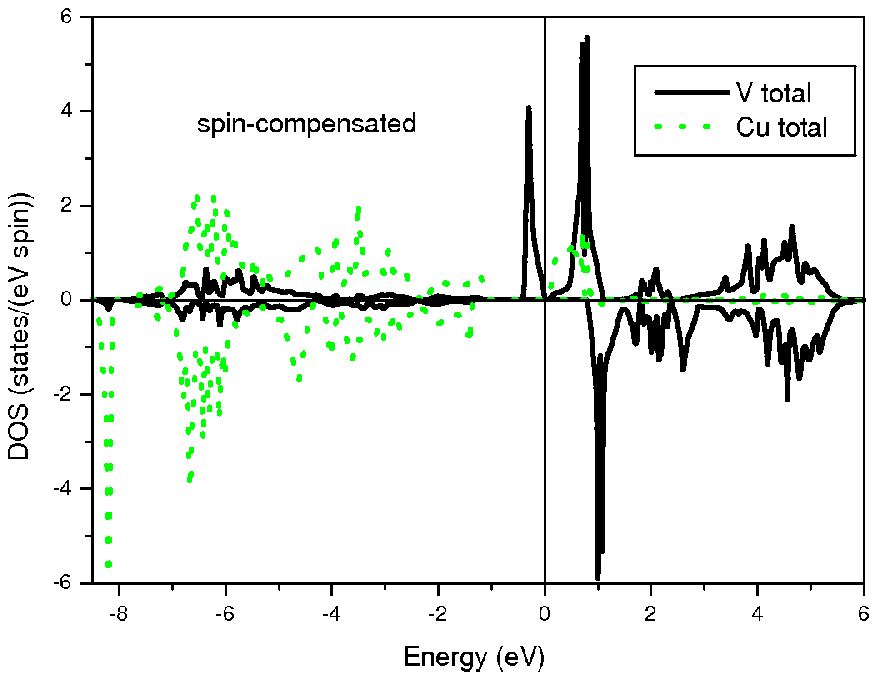}
\includegraphics[width=\columnwidth,draft=false]{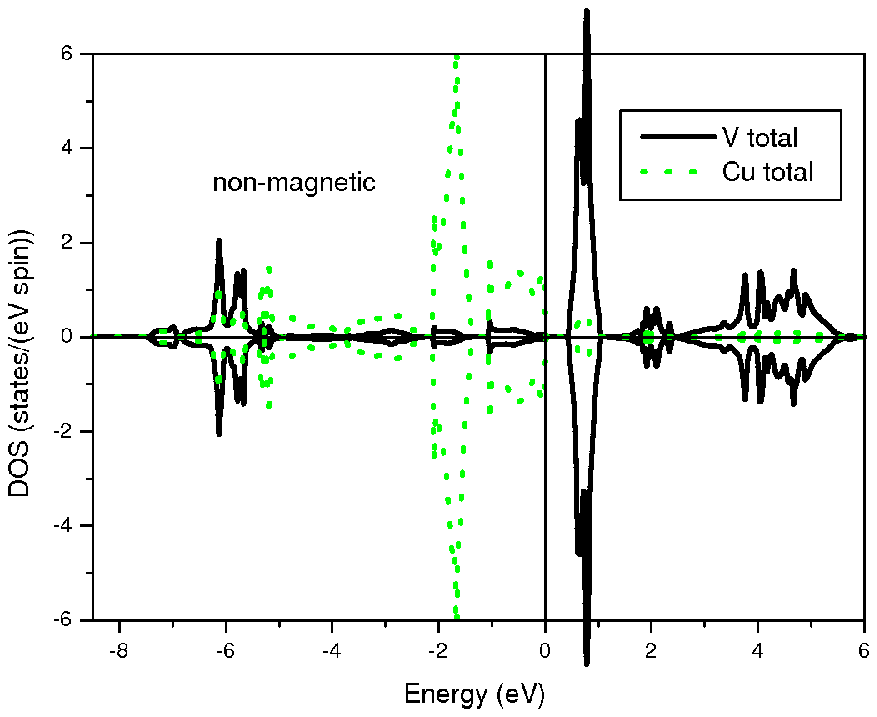}
\caption{Density of states (DOS) plots of the spin-compensated (top) and non-magnetic (bottom) solutions. Upper (lower) panels show the majority (minority) spin density of states, in each case. A half-metallic solution is found for certain values of U for the spin-compensated solution, as shown in the top figure for U= 6 eV for Cu and 3.5 eV for V. An insulating solution would occur for larger values of U (close to the unrealistic limit). The bottom figure shows the insulating, full-shell solution found to be the ground state non-magnetic solution, in this case showing the DOS for U= 6 eV for Cu and 3.5 eV for V.}\label{dos}
\end{center}
\end{figure}

Figure \ref{dos} shows the density of states (DOS) of the two charge configurations found. The spin-compensated solution (upper panel) is a half-metallic semimetal, with non-zero DOS at the Fermi level for the majority spin channel, whereas a wide gap appears in the minority-spin channel. For larger values of U, a gap opens also in the majority spin channel producing a compensated ferrimagnetic insulator.\cite{vpardo_hmaf} However, in the reasonable range of U values we present here (6 eV for Cu and 3.5 eV for V), the system is half-metallic and also spin-compensated. The value of the moments, both for Cu and V, is slightly over 0.7 $\mu_B$ inside the muffin-tin spheres considered, a value that is substantially larger than those presented in Ref. \onlinecite{la2vcuo6_guo} (0.54 $\mu_B$ for V and 0.4 $\mu_B$ for Cu inside the muffin-tin spheres) and large enough for the S=1/2 d$^1$-d$^9$ description to be realistic.

The spin-compensated solution shows (see Fig. \ref{dos}, upper panel) that around the Fermi level, the V t$_{2g}$ triplet is split by the symmetry lowering, and the lower d$_{xy}$ level becomes fully occupied. Just above the Fermi level lies the \{$d_{xz},d_{yz}$\} doublet. Also, the Cu d bands can be easily described: around the Fermi level there is one almost unoccupied e$_g$ band (of $x^2$-$y^2$ symmetry). Its minority-spin counterpart is located 8 eV below the Fermi level, moved to lower energy by the splitting caused by U (6 eV for Cu).

\begin{figure*}[ht]
\begin{center}
\includegraphics[width=\columnwidth,draft=false]{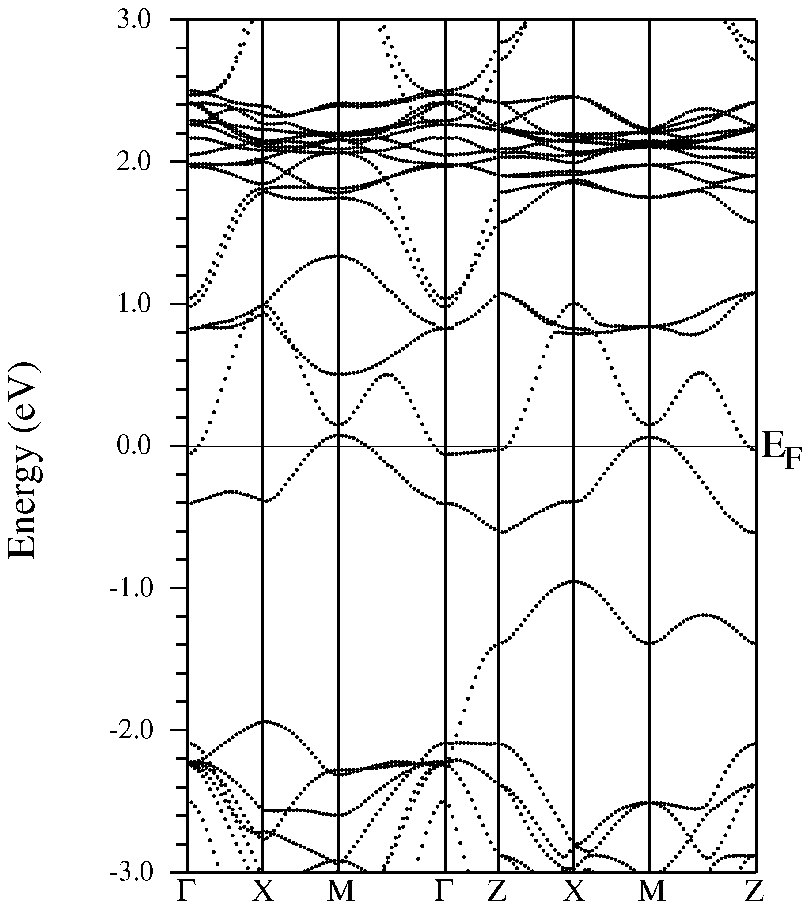}
\includegraphics[width=\columnwidth,draft=false]{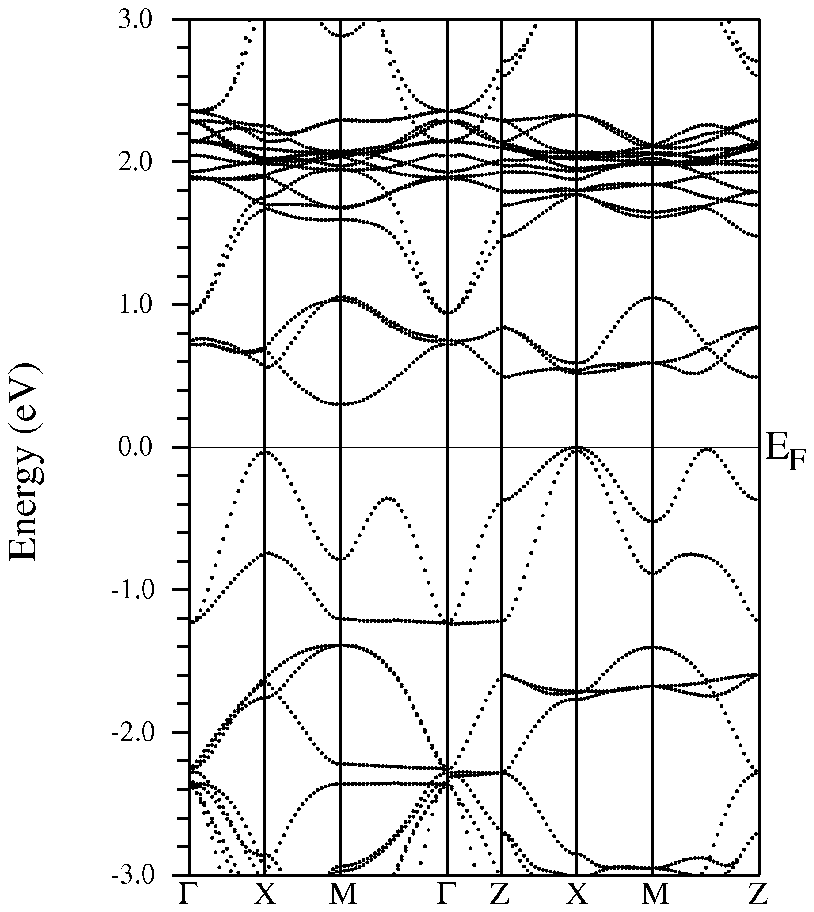}
\caption{Band structures of the two states that are analyzed in the text. On the left side, the majority spin of the spin-compensated solution is shown. Two bands, a nearly filled V d$_{xy}$ valence band and an almost unoccupied Cu d$_{x^2-y^2}$ conduction band, overlap, leading to a half-metallic semimetal with very small Fermi surfaces. On the right side, the majority spin (both channels are equivalent) of the non-magnetic state is pictured, consisting of filled Cu $3d$ bands and empty V $3d$ bands. For both Cu and V the $e_g$-$t_{2g}$ splitting is evident.  The flat bands at 2 eV are La $4f$ bands; O $2p$ states lie below -2 eV.}\label{bs}
\end{center}
\end{figure*}

Figure \ref{bs} displays the band structures for the two states we are discussing. For analyzing the electronic structure of the spin-compensated solution, it is interesting to study only the spin-up band structure, since the spin-down bands shows an uninteresting gap around the Fermi level, as we saw above when discussing its DOS plot. The left panel of Fig. \ref{bs} shows two bands crossing the Fermi level. There is an almost completely full V d$_{xy}$ band that forms a hole pocket around M, and also the Cu d$_{x^2-y^2}$ crosses the Fermi level near $\Gamma$, being nearly unoccupied. It is clear from the band structure plot that a slight increase in the value of U utilized for Cu and/or V would make the system be an insulator, as was found previously\cite{vpardo_hmaf} and described in detail above. The right panel of Fig. \ref{bs} shows the band structure of the non-magnetic solution. Just below the gap two occupied Cu e$_g$ bands can be identified, crystal-field split with respect to the Cu t$_{2g}$ bands, the triplet around -2 eV. Just above the gap, the three V t$_{2g}$ bands can also be identified. They are crystal-field split with the two e$_{g}$ bands only slightly higher in energy. The gap will tend to increase for larger values of U, but already an insulating phase occurs at small values of U.

This spin-compensated solution is however not the ground state according to the calculated total energy, both in the GGA (U= 0 level) limit and also when the LDA+U method is utilized with moderate values of U (see below). The ground state is an unexpected non-magnetic ionic insulator, characterized by a d$^0$-d$^{10}$ spherically symmetric empty shell/full shell combination. This non-magnetic state is insulating for a broad range of values of U, including the U=0 GGA limit. For the values of U we think are reasonable (U= 6 eV for Cu and 3.5 eV for V), the non-magnetic solution is more stable than the spin-compensated one by about 220 meV per metal atom. The stability of that solution gets greater as strong correlation effects (quantified by U in our calculations) are reduced. In GGA (U= 0), the spin-compensated solution cannot be obtained, as we have mentioned above. However, at larger U values (starting at U= 7 eV for Cu and 4 eV for V), the magnetic but spin-compensated solution
becomes insulating.  It also becomes more stable than the non-magnetic solution, which is always insulating. 
Thus, for the whole U spectrum, an insulating solution is always the most favorable energetically, ruling out the possibility of this system being a candidate for showing half-metallicity in the absence of a total magnetic moment.

We have also tried a different symmetry that could lead to new solutions:\cite{la2vcuo6_guo} a layered structure with alternating V  and Cu layers perpendicular to the $z$-axis.  This structure leads to ferromagnetic coupling within the V plane, produced by the distortion of the octahedral cage around V that leads to one electron occupying the \{$d_{xz},d_{yz}$\} doublet. The coupling in the Cu plane is AF. 
The total energy of such solution is however less stable than the non-magnetic ground state by about 10 meV/metal for U= 0, increasing for larger values of U.

One may wonder about the contributions of the differing formal
valences to the total energy balance, for the two states that
compete for the ground state in this compound.  The more highly
charge differentiated V$^{5+}$-Cu$^{+}$ state is lower in energy than 
the V$^{4+}$-Cu$^{2+}$ state (except for large $U$), whereas a classical picture would 
lead to a considerably higher electrostatic energy that could of course be
compensated somewhat by other contributions.  We just want 
to point out, as has been noted several times 
previously,\cite{Weht,KWL,KWL2} that 
while the formal charge state of ions in oxides is an extremely
important and useful concept and label, the actual charge difference
can be small.  For these two V$^{5+}$-Cu$^{+}$ and V$^{4+}$-Cu$^{2+}$
configurations, the respective actual integrated $3d$ charges within
the LAPW inscribed spheres are, for V, 2.00 vs. 2.10, and for Cu,
8.98 vs. 8.71.  Most interesting here is the very small difference
for V, only 0.10 electron, and the fact that judged on actual
charge, the ionic character seems to be more representative of a 
3+ ion in {\it both} V$^{5+}$ and
V$^{4+}$ formal oxidation states.  Another point of mentioning this is that
the small actual charge difference makes more plausible the small energy
difference that we find.

Another unusual aspect is that
this charge has,  from Fig. 1, substantial contribution to 
hybridization contributions at -5 to -6 eV (with respect to the
Fermi level).  The formal valence nevertheless describes the 
overall electronic structure very well (``fatbands'' plots of
wavefunction character show that all five V bands are unoccupied
for the V$^{5+}$ state), and the spin moments can only be
understood and discussed using the formal charge states. The 
best understanding of these effects are based on hybridized
(Wannier) orbitals,\cite{Weht} which include the effects of 
chemical bonding.

\section{Summary}

In summary, we have identified the possible electronic phases of La$_2$VCuO$_6$, one of them showing spin-compensated half-metallicity. However, the ground state at low and moderate values of U according to our total energy calculations is an unexpected non-magnetic insulating state, that comes about with the unexpected combination of valencies Cu$^+$ and V$^{5+}$. At large values of U, a spin-compensated solution is more stable, but showing semimetallic to insulating character depending on the strength of on-site Coulomb repulsion. We suggest that this compound would be an interesting double perovskite system to study experimentally, to elucidate which of the two phases occurs and whether a metallic phase exists in a narrow window of parameter space.

\acknowledgments
This project was supported by DOE grant DE-FG02-04ER46111.


\begin{thebibliography}{39}
\expandafter\ifx\csname natexlab\endcsname\relax\def\natexlab#1{#1}\fi
\expandafter\ifx\csname bibnamefont\endcsname\relax
  \def\bibnamefont#1{#1}\fi
\expandafter\ifx\csname bibfnamefont\endcsname\relax
  \def\bibfnamefont#1{#1}\fi
\expandafter\ifx\csname citenamefont\endcsname\relax
  \def\citenamefont#1{#1}\fi
\expandafter\ifx\csname url\endcsname\relax
  \def\url#1{\texttt{#1}}\fi
\expandafter\ifx\csname urlprefix\endcsname\relax\def\urlprefix{URL }\fi
\providecommand{\bibinfo}[2]{#2}
\providecommand{\eprint}[2][]{\url{#2}}

\bibitem[{\citenamefont{Kobayashi et~al.}(1998)\citenamefont{Kobayashi, Kimura,
  Sawada, Terakura, and Tokura}}]{sr2femoo6}
\bibinfo{author}{\bibfnamefont{K.~I.} \bibnamefont{Kobayashi}},
  \bibinfo{author}{\bibfnamefont{T.}~\bibnamefont{Kimura}},
  \bibinfo{author}{\bibfnamefont{H.}~\bibnamefont{Sawada}},
  \bibinfo{author}{\bibfnamefont{K.}~\bibnamefont{Terakura}}, \bibnamefont{and}
  \bibinfo{author}{\bibfnamefont{Y.}~\bibnamefont{Tokura}},
  \bibinfo{journal}{Nature} \textbf{\bibinfo{volume}{395}},
  \bibinfo{pages}{677} (\bibinfo{year}{1998}).

\bibitem[{\citenamefont{Katsnelson et~al.}(2008)\citenamefont{Katsnelson,
  Irkhin, Chioncel, Lichtenstein, and Groot}}]{katsnelsonRMP}
\bibinfo{author}{\bibfnamefont{M.~I.} \bibnamefont{Katsnelson}},
  \bibinfo{author}{\bibfnamefont{V.~Y.} \bibnamefont{Irkhin}},
  \bibinfo{author}{\bibfnamefont{L.}~\bibnamefont{Chioncel}},
  \bibinfo{author}{\bibfnamefont{A.~I.} \bibnamefont{Lichtenstein}},
  \bibnamefont{and} \bibinfo{author}{\bibfnamefont{R.~A.} \bibnamefont{Groot}},
  \bibinfo{journal}{Rev. Mod. Phys.} \textbf{\bibinfo{volume}{80}},
  \bibinfo{pages}{315} (\bibinfo{year}{2008}).

\bibitem[{\citenamefont{Spaldin and Pickett}(2003)}]{hmafm_review_jssc}
\bibinfo{author}{\bibfnamefont{N.~A.} \bibnamefont{Spaldin}} \bibnamefont{and}
  \bibinfo{author}{\bibfnamefont{W.~E.} \bibnamefont{Pickett}},
  \bibinfo{journal}{J. Solid State Chem.} \textbf{\bibinfo{volume}{176}},
  \bibinfo{pages}{615} (\bibinfo{year}{2003}).

\bibitem[{\citenamefont{Felser et~al.}(2007)\citenamefont{Felser, Fecher, and
  Balke}}]{hmafm_review_ange}
\bibinfo{author}{\bibfnamefont{C.}~\bibnamefont{Felser}},
  \bibinfo{author}{\bibfnamefont{G.~H.} \bibnamefont{Fecher}},
  \bibnamefont{and} \bibinfo{author}{\bibfnamefont{B.}~\bibnamefont{Balke}},
  \bibinfo{journal}{Angew. Chem. Int. Ed.} \textbf{\bibinfo{volume}{46}},
  \bibinfo{pages}{668} (\bibinfo{year}{2007}).

\bibitem[{\citenamefont{de~Groot et~al.}(1983)\citenamefont{de~Groot, Mueller,
  van Engen, and Buschow}}]{degroot}
\bibinfo{author}{\bibfnamefont{R.~A.} \bibnamefont{de~Groot}},
  \bibinfo{author}{\bibfnamefont{F.~M.} \bibnamefont{Mueller}},
  \bibinfo{author}{\bibfnamefont{P.~G.} \bibnamefont{van Engen}},
  \bibnamefont{and} \bibinfo{author}{\bibfnamefont{K.~H.~J.}
  \bibnamefont{Buschow}}, \bibinfo{journal}{Phys. Rev. Lett.}
  \textbf{\bibinfo{volume}{50}}, \bibinfo{pages}{2024} (\bibinfo{year}{1983}).

\bibitem[{\citenamefont{Schwarz}(1986)}]{cro2_schwarz}
\bibinfo{author}{\bibfnamefont{K.}~\bibnamefont{Schwarz}}, \bibinfo{journal}{J.
  Phys. F: Met. Phys.} \textbf{\bibinfo{volume}{16}}, \bibinfo{pages}{L211}
  (\bibinfo{year}{1986}).

\bibitem[{\citenamefont{Pickett}(1998)}]{pickett_hmafm}
\bibinfo{author}{\bibfnamefont{W.~E.} \bibnamefont{Pickett}},
  \bibinfo{journal}{Phys. Rev. B} \textbf{\bibinfo{volume}{57}},
  \bibinfo{pages}{10613} (\bibinfo{year}{1998}).

\bibitem[{\citenamefont{Pardo and Pickett}(2009)}]{vpardo_hmaf}
\bibinfo{author}{\bibfnamefont{V.}~\bibnamefont{Pardo}} \bibnamefont{and}
  \bibinfo{author}{\bibfnamefont{W.~E.} \bibnamefont{Pickett}},
  \bibinfo{journal}{Phys. Rev. B} \textbf{\bibinfo{volume}{80}},
  \bibinfo{pages}{054415} (\bibinfo{year}{2009}).

\bibitem[{\citenamefont{Wang et~al.}(2009)\citenamefont{Wang, Lee, and
  Guo}}]{la2vcuo6_guo}
\bibinfo{author}{\bibfnamefont{Y.~K.} \bibnamefont{Wang}},
  \bibinfo{author}{\bibfnamefont{P.~H.} \bibnamefont{Lee}}, \bibnamefont{and}
  \bibinfo{author}{\bibfnamefont{G.~Y.} \bibnamefont{Guo}},
  \bibinfo{journal}{Phys. Rev. B} \textbf{\bibinfo{volume}{80}},
  \bibinfo{pages}{224418} (\bibinfo{year}{2009}).

\bibitem[{\citenamefont{Androulakis et~al.}(2002)\citenamefont{Androulakis,
  Katsakaris, and Gianpintzakis}}]{androulakis}
\bibinfo{author}{\bibfnamefont{J.}~\bibnamefont{Androulakis}},
  \bibinfo{author}{\bibfnamefont{N.}~\bibnamefont{Katsakaris}},
  \bibnamefont{and}
  \bibinfo{author}{\bibfnamefont{J.}~\bibnamefont{Gianpintzakis}},
  \bibinfo{journal}{Solid State Commun.} \textbf{\bibinfo{volume}{124}},
  \bibinfo{pages}{77} (\bibinfo{year}{2002}).

\bibitem[{\citenamefont{Park et~al.}(2002)\citenamefont{Park, Kwon, and
  Min}}]{lasrvruo6}
\bibinfo{author}{\bibfnamefont{J.~H.} \bibnamefont{Park}},
  \bibinfo{author}{\bibfnamefont{S.~K.} \bibnamefont{Kwon}}, \bibnamefont{and}
  \bibinfo{author}{\bibfnamefont{B.~I.} \bibnamefont{Min}},
  \bibinfo{journal}{Phys. Rev. B} \textbf{\bibinfo{volume}{65}},
  \bibinfo{pages}{174401} (\bibinfo{year}{2002}).

\bibitem[{\citenamefont{Park and Min}(2005)}]{lacavmoo6}
\bibinfo{author}{\bibfnamefont{M.~S.} \bibnamefont{Park}} \bibnamefont{and}
  \bibinfo{author}{\bibfnamefont{B.~I.} \bibnamefont{Min}},
  \bibinfo{journal}{Phys. Rev. B} \textbf{\bibinfo{volume}{71}},
  \bibinfo{pages}{052405} (\bibinfo{year}{2005}).

\bibitem[{\citenamefont{Wang and Guo}(2006)}]{lacavoso6}
\bibinfo{author}{\bibfnamefont{Y.~K.} \bibnamefont{Wang}} \bibnamefont{and}
  \bibinfo{author}{\bibfnamefont{G.~Y.} \bibnamefont{Guo}},
  \bibinfo{journal}{Phys. Rev. B} \textbf{\bibinfo{volume}{73}},
  \bibinfo{pages}{064424} (\bibinfo{year}{2006}).

\bibitem[{\citenamefont{Wang et~al.}(2010)\citenamefont{Wang, Song, and
  Wu}}]{hm_cpl}
\bibinfo{author}{\bibfnamefont{J.}~\bibnamefont{Wang}},
  \bibinfo{author}{\bibfnamefont{W.~Y.} \bibnamefont{Song}}, \bibnamefont{and}
  \bibinfo{author}{\bibfnamefont{Z.~J.} \bibnamefont{Wu}},
  \bibinfo{journal}{Chem. Phys. Lett.} \textbf{\bibinfo{volume}{492}},
  \bibinfo{pages}{241} (\bibinfo{year}{2010}).

\bibitem[{\citenamefont{Nie and Hu}(2008)}]{hmafm_per}
\bibinfo{author}{\bibfnamefont{Y.~M.} \bibnamefont{Nie}} \bibnamefont{and}
  \bibinfo{author}{\bibfnamefont{X.}~\bibnamefont{Hu}}, \bibinfo{journal}{Phys.
  Rev. Lett.} \textbf{\bibinfo{volume}{100}}, \bibinfo{pages}{117203}
  (\bibinfo{year}{2008}).

\bibitem[{\citenamefont{Jana et~al.}(2010)\citenamefont{Jana, Singh, Kaushik,
  Meneghini, Pal, Knut, Karis, Dasgupta, Siruguri, and Ray}}]{hmafm_dp_prb}
\bibinfo{author}{\bibfnamefont{S.}~\bibnamefont{Jana}},
  \bibinfo{author}{\bibfnamefont{V.}~\bibnamefont{Singh}},
  \bibinfo{author}{\bibfnamefont{S.~D.} \bibnamefont{Kaushik}},
  \bibinfo{author}{\bibfnamefont{C.}~\bibnamefont{Meneghini}},
  \bibinfo{author}{\bibfnamefont{P.}~\bibnamefont{Pal}},
  \bibinfo{author}{\bibfnamefont{R.}~\bibnamefont{Knut}},
  \bibinfo{author}{\bibfnamefont{O.}~\bibnamefont{Karis}},
  \bibinfo{author}{\bibfnamefont{I.}~\bibnamefont{Dasgupta}},
  \bibinfo{author}{\bibfnamefont{V.}~\bibnamefont{Siruguri}}, \bibnamefont{and}
  \bibinfo{author}{\bibfnamefont{S.}~\bibnamefont{Ray}},
  \bibinfo{journal}{Phys. Rev. B} \textbf{\bibinfo{volume}{82}},
  \bibinfo{pages}{180407} (\bibinfo{year}{2010}).

\bibitem[{\citenamefont{Uehara et~al.}(2004)\citenamefont{Uehara, Yamada, and
  Kimishima}}]{hmafm_dp_ssc}
\bibinfo{author}{\bibfnamefont{M.}~\bibnamefont{Uehara}},
  \bibinfo{author}{\bibfnamefont{M.}~\bibnamefont{Yamada}}, \bibnamefont{and}
  \bibinfo{author}{\bibfnamefont{Y.}~\bibnamefont{Kimishima}},
  \bibinfo{journal}{Solid State Commun.} \textbf{\bibinfo{volume}{129}},
  \bibinfo{pages}{385} (\bibinfo{year}{2004}).

\bibitem[{\citenamefont{Chen et~al.}(2011)\citenamefont{Chen, Xiao, Liu, Lee,
  and Wang}}]{hmafm_dp_chen_jmmm}
\bibinfo{author}{\bibfnamefont{S.~H.} \bibnamefont{Chen}},
  \bibinfo{author}{\bibfnamefont{Z.~R.} \bibnamefont{Xiao}},
  \bibinfo{author}{\bibfnamefont{Y.~P.} \bibnamefont{Liu}},
  \bibinfo{author}{\bibfnamefont{P.~H.} \bibnamefont{Lee}}, \bibnamefont{and}
  \bibinfo{author}{\bibfnamefont{Y.~K.} \bibnamefont{Wang}},
  \bibinfo{journal}{J. Magn. Magn. Mater.} \textbf{\bibinfo{volume}{323}},
  \bibinfo{pages}{176} (\bibinfo{year}{2011}).

\bibitem[{\citenamefont{Chen et~al.}(2010)\citenamefont{Chen, Xiao, Liu, and
  Wang}}]{hmafm_dp_chen_jap}
\bibinfo{author}{\bibfnamefont{S.~H.} \bibnamefont{Chen}},
  \bibinfo{author}{\bibfnamefont{Z.~R.} \bibnamefont{Xiao}},
  \bibinfo{author}{\bibfnamefont{Y.~P.} \bibnamefont{Liu}}, \bibnamefont{and}
  \bibinfo{author}{\bibfnamefont{Y.~K.} \bibnamefont{Wang}},
  \bibinfo{journal}{J. Appl. Phys.} \textbf{\bibinfo{volume}{108}},
  \bibinfo{pages}{093908} (\bibinfo{year}{2010}).

\bibitem[{\citenamefont{Nakao}(2008)}]{hmafm_ml_physb}
\bibinfo{author}{\bibfnamefont{M.}~\bibnamefont{Nakao}},
  \bibinfo{journal}{Physica B} \textbf{\bibinfo{volume}{403}},
  \bibinfo{pages}{1431} (\bibinfo{year}{2008}).

\bibitem[{\citenamefont{Nakao}(2007)}]{hmafm_ml_jmmm}
\bibinfo{author}{\bibfnamefont{M.}~\bibnamefont{Nakao}}, \bibinfo{journal}{J.
  Magn. Magn. Mater.} \textbf{\bibinfo{volume}{310}}, \bibinfo{pages}{2259}
  (\bibinfo{year}{2007}).

\bibitem[{\citenamefont{Hu and Hu}(2010)}]{hmafm_pnic_jpcc}
\bibinfo{author}{\bibfnamefont{S.~J.} \bibnamefont{Hu}} \bibnamefont{and}
  \bibinfo{author}{\bibfnamefont{X.}~\bibnamefont{Hu}}, \bibinfo{journal}{J.
  Phys. Chem. C} \textbf{\bibinfo{volume}{114}}, \bibinfo{pages}{11614}
  (\bibinfo{year}{2010}).

\bibitem[{\citenamefont{Lee et~al.}(2009)\citenamefont{Lee, Bialek, and
  Kim}}]{hmafm_pnic_jap}
\bibinfo{author}{\bibfnamefont{J.~I.} \bibnamefont{Lee}},
  \bibinfo{author}{\bibfnamefont{B.}~\bibnamefont{Bialek}}, \bibnamefont{and}
  \bibinfo{author}{\bibfnamefont{M.}~\bibnamefont{Kim}}, \bibinfo{journal}{J.
  Appl. Phys.} \textbf{\bibinfo{volume}{105}}, \bibinfo{pages}{07E508}
  (\bibinfo{year}{2009}).

\bibitem[{\citenamefont{Long et~al.}(2009)\citenamefont{Long, Ogura, and
  Akai}}]{hmafm_pnic_jpcm}
\bibinfo{author}{\bibfnamefont{N.~H.} \bibnamefont{Long}},
  \bibinfo{author}{\bibfnamefont{M.}~\bibnamefont{Ogura}}, \bibnamefont{and}
  \bibinfo{author}{\bibfnamefont{H.}~\bibnamefont{Akai}}, \bibinfo{journal}{J.
  Phys.: Condens. Matter} \textbf{\bibinfo{volume}{21}},
  \bibinfo{pages}{064241} (\bibinfo{year}{2009}).

\bibitem[{\citenamefont{Endo et~al.}(1995)\citenamefont{Endo, Matsuda, Ooiwa,
  and Itoh}}]{hmafm_heus_jpsj}
\bibinfo{author}{\bibfnamefont{K.}~\bibnamefont{Endo}},
  \bibinfo{author}{\bibfnamefont{H.}~\bibnamefont{Matsuda}},
  \bibinfo{author}{\bibfnamefont{K.}~\bibnamefont{Ooiwa}}, \bibnamefont{and}
  \bibinfo{author}{\bibfnamefont{K.}~\bibnamefont{Itoh}}, \bibinfo{journal}{J.
  Phys. Soc. Japan} \textbf{\bibinfo{volume}{64}}, \bibinfo{pages}{2329}
  (\bibinfo{year}{1995}).

\bibitem[{\citenamefont{Wurmehl et~al.}(2006)\citenamefont{Wurmehl, Kandpal,
  Fecher, and Felser}}]{hmafm_heus_jpcm}
\bibinfo{author}{\bibfnamefont{S.}~\bibnamefont{Wurmehl}},
  \bibinfo{author}{\bibfnamefont{H.~C.} \bibnamefont{Kandpal}},
  \bibinfo{author}{\bibfnamefont{G.~H.} \bibnamefont{Fecher}},
  \bibnamefont{and} \bibinfo{author}{\bibfnamefont{C.}~\bibnamefont{Felser}},
  \bibinfo{journal}{J. Phys.: Condens. Matter} \textbf{\bibinfo{volume}{18}},
  \bibinfo{pages}{6171} (\bibinfo{year}{2006}).

\bibitem[{\citenamefont{Ozdogan}(2009)}]{hmafm_heus_jmmm}
\bibinfo{author}{\bibfnamefont{K.}~\bibnamefont{Ozdogan}}, \bibinfo{journal}{J.
  Magn. Magn. Mater.} \textbf{\bibinfo{volume}{321}}, \bibinfo{pages}{L34}
  (\bibinfo{year}{2009}).

\bibitem[{\citenamefont{Galanakis et~al.}(2002)\citenamefont{Galanakis,
  Dederichs, and Papanikolaou}}]{hmafm_heus_prb}
\bibinfo{author}{\bibfnamefont{I.}~\bibnamefont{Galanakis}},
  \bibinfo{author}{\bibfnamefont{P.~H.} \bibnamefont{Dederichs}},
  \bibnamefont{and}
  \bibinfo{author}{\bibfnamefont{N.}~\bibnamefont{Papanikolaou}},
  \bibinfo{journal}{Phys. Rev. B} \textbf{\bibinfo{volume}{66}},
  \bibinfo{pages}{134428} (\bibinfo{year}{2002}).

\bibitem[{\citenamefont{Bergqvist and Dederichs}(2007)}]{hmafm_dms}
\bibinfo{author}{\bibfnamefont{L.}~\bibnamefont{Bergqvist}} \bibnamefont{and}
  \bibinfo{author}{\bibfnamefont{P.~H.} \bibnamefont{Dederichs}},
  \bibinfo{journal}{J. Phys.: Condens. Matter} \textbf{\bibinfo{volume}{19}},
  \bibinfo{pages}{216220} (\bibinfo{year}{2007}).

\bibitem[{\citenamefont{Rudd and Pickett}(1998)}]{pickett_sss}
\bibinfo{author}{\bibfnamefont{R.~E.} \bibnamefont{Rudd}} \bibnamefont{and}
  \bibinfo{author}{\bibfnamefont{W.~E.} \bibnamefont{Pickett}},
  \bibinfo{journal}{Phys. Rev. B} \textbf{\bibinfo{volume}{57}},
  \bibinfo{pages}{557} (\bibinfo{year}{1998}).

\bibitem[{\citenamefont{Hohenberg and Kohn}(1964)}]{dft}
\bibinfo{author}{\bibfnamefont{P.}~\bibnamefont{Hohenberg}} \bibnamefont{and}
  \bibinfo{author}{\bibfnamefont{W.}~\bibnamefont{Kohn}},
  \bibinfo{journal}{Phys. Rev.} \textbf{\bibinfo{volume}{136}},
  \bibinfo{pages}{B864} (\bibinfo{year}{1964}).

\bibitem[{\citenamefont{Schwarz and Blaha}(2003)}]{wien}
\bibinfo{author}{\bibfnamefont{K.}~\bibnamefont{Schwarz}} \bibnamefont{and}
  \bibinfo{author}{\bibfnamefont{P.}~\bibnamefont{Blaha}},
  \bibinfo{journal}{Comp. Mat. Sci.} \textbf{\bibinfo{volume}{28}},
  \bibinfo{pages}{259} (\bibinfo{year}{2003}).

\bibitem[{\citenamefont{Sj{\"o}stedt et~al.}(2000)\citenamefont{Sj{\"o}stedt,
  N{\"o}rdstrom, and Singh}}]{sjo}
\bibinfo{author}{\bibfnamefont{E.}~\bibnamefont{Sj{\"o}stedt}},
  \bibinfo{author}{\bibfnamefont{L.}~\bibnamefont{N{\"o}rdstrom}},
  \bibnamefont{and} \bibinfo{author}{\bibfnamefont{D.~J.} \bibnamefont{Singh}},
  \bibinfo{journal}{Solid State Commun.} \textbf{\bibinfo{volume}{114}},
  \bibinfo{pages}{15} (\bibinfo{year}{2000}).

\bibitem[{\citenamefont{Perdew et~al.}(1996)\citenamefont{Perdew, Burke, and
  Ernzerhof}}]{gga}
\bibinfo{author}{\bibfnamefont{J.~P.} \bibnamefont{Perdew}},
  \bibinfo{author}{\bibfnamefont{K.}~\bibnamefont{Burke}}, \bibnamefont{and}
  \bibinfo{author}{\bibfnamefont{M.}~\bibnamefont{Ernzerhof}},
  \bibinfo{journal}{Phys.\ Rev. Lett.} \textbf{\bibinfo{volume}{77}},
  \bibinfo{pages}{3865} (\bibinfo{year}{1996}).

\bibitem[{\citenamefont{Anisimov et~al.}(1991)\citenamefont{Anisimov, Zaanen,
  and Andersen}}]{sic1}
\bibinfo{author}{\bibfnamefont{V.~I.} \bibnamefont{Anisimov}},
  \bibinfo{author}{\bibfnamefont{J.}~\bibnamefont{Zaanen}}, \bibnamefont{and}
  \bibinfo{author}{\bibfnamefont{O.~K.} \bibnamefont{Andersen}},
  \bibinfo{journal}{Phys. Rev. B} \textbf{\bibinfo{volume}{44}},
  \bibinfo{pages}{943} (\bibinfo{year}{1991}).

\bibitem[{\citenamefont{Ylvisaker et~al.}(2009)\citenamefont{Ylvisaker,
  Pickett, and Koepernik}}]{sic2}
\bibinfo{author}{\bibfnamefont{E.~R.} \bibnamefont{Ylvisaker}},
  \bibinfo{author}{\bibfnamefont{W.~E.} \bibnamefont{Pickett}},
  \bibnamefont{and}
  \bibinfo{author}{\bibfnamefont{K.}~\bibnamefont{Koepernik}},
  \bibinfo{journal}{Phys. Rev. B} \textbf{\bibinfo{volume}{79}},
  \bibinfo{pages}{035103} (\bibinfo{year}{2009}).

\bibitem[{\citenamefont{Weht and Pickett}(1998)}]{Weht}
\bibinfo{author}{\bibfnamefont{R.}~\bibnamefont{Weht}} \bibnamefont{and}
  \bibinfo{author}{\bibfnamefont{W.~E.} \bibnamefont{Pickett}},
  \bibinfo{journal}{Phys. Rev. Lett.} \textbf{\bibinfo{volume}{81}},
  \bibinfo{pages}{2502} (\bibinfo{year}{1998}).

\bibitem[{\citenamefont{Lee and Pickett}(2004)}]{KWL}
\bibinfo{author}{\bibfnamefont{K.~W.} \bibnamefont{Lee}} \bibnamefont{and}
  \bibinfo{author}{\bibfnamefont{W.~E.} \bibnamefont{Pickett}},
  \bibinfo{journal}{Phys. Rev. B} \textbf{\bibinfo{volume}{70}},
  \bibinfo{pages}{165109} (\bibinfo{year}{2004}).

\bibitem[{\citenamefont{Lee et~al.}(2004)\citenamefont{Lee, Kune\v{s}, and
  Pickett}}]{KWL2}
\bibinfo{author}{\bibfnamefont{K.~W.} \bibnamefont{Lee}},
  \bibinfo{author}{\bibfnamefont{J.}~\bibnamefont{Kune\v{s}}},
  \bibnamefont{and} \bibinfo{author}{\bibfnamefont{W.~E.}
  \bibnamefont{Pickett}}, \bibinfo{journal}{Phys. Rev. B}
  \textbf{\bibinfo{volume}{70}}, \bibinfo{pages}{045104}
  (\bibinfo{year}{2004}).

\end{thebibliography}

\end{document}